\documentclass[twocolumn,prb,floats,showpacs,superscriptaddress]{revtex4}
\usepackage{graphicx,epsfig}
\usepackage{times}
\usepackage{graphics,dcolumn,bm,float}
\usepackage{amssymb,amsmath,rotate,color}

\begin{document}
\unitlength 1 cm
\newcommand{\be}{\begin{equation}}
\newcommand{\ee}{\end{equation}}
\newcommand{\bearr}{\begin{eqnarray}}
\newcommand{\eearr}{\end{eqnarray}}
\newcommand{\nn}{\nonumber}
\newcommand{\vk}{\vec k}
\newcommand{\vp}{\vec p}
\newcommand{\vq}{\vec q}
\newcommand{\vkp}{\vec {k'}}
\newcommand{\vpp}{\vec {p'}}
\newcommand{\vqp}{\vec {q'}}
\newcommand{\bk}{{\bf k}}
\newcommand{\bp}{{\bf p}}
\newcommand{\bq}{{\bf q}}
\newcommand{\br}{{\bf r}}
\newcommand{\bR}{{\bf R}}
\newcommand{\up}{\uparrow}
\newcommand{\down}{\downarrow}
\newcommand{\fns}{\footnotesize}
\newcommand{\ns}{\normalsize}
\newcommand{\cdag}{c^{\dagger}}

\title{Magnetic edge states in Aharonov-Bohm graphene quantum rings}
\author{R. Farghadan}\email{rfarghadan@kashanu.ac.ir}
\affiliation{Department of Physics, University of Kashan, Kashan, Iran }
\author{A. Saffarzadeh}
\affiliation{Department of Physics, Payame Noor University, P.O.
Box 19395-3697 Tehran, Iran} \affiliation{Department of Physics,
Simon Fraser University, Burnaby, British Columbia, Canada V5A
1S6}
\author{E. Heidari Semiromi}
\affiliation{Department of Physics, University of Kashan, Kashan, Iran }
\date{\today}

\begin{abstract}
The effect of electron-electron interaction on the electronic
structure of Aharonov-Bohm (AB) graphene quantum rings (GQRs) is
explored theoretically using the single-band tight-binding
Hamiltonian and the mean-field Hubbard model. The electronic
states and magnetic properties of hexagonal, triangular and
circular GQRs with different sizes and zigzag edge terminations
are studied. The results show that, although the AB oscillations
in the all types of nanoring are affected by the interaction, the
spin splitting in the AB oscillations strongly depends on the
geometry and the size of graphene nanorings. We found that the
total spin of hexagonal and circular rings is zero and therefore,
no spin splitting can be observed in the AB oscillations. However,
the non-zero magnetization of the triangular rings breaks the
degeneracy between spin-up and spin-down electrons, which produces
spin-polarized AB oscillations.
\end{abstract}


\maketitle
\section{Introduction}
In  nanoscience and nanotechnology, quantum ring nanostructures
with phase-coherence phenomena, such as persistent currents and
the Aharonov-Bohm (AB) oscillations, are considered as good
candidates for quantum interference devices due to their unique
topology \cite{Kleemans,Kot,Guclu}. The existence of persistent
current, a direct consequence of the AB effect \cite{Aharonov1959}
in normal electronic mesoscopic rings pierced by a magnetic flux,
was observed both theoretically \cite{Buttiker1983} and
experimentally \cite{Webb1985} by pointing out that the flux
quantum $\phi_0=hc/e$ determines the period of the current.
Moreover, the AB effect in metal rings \cite{Webb1985} and the
flux dependence of persistent current in mesoscopic rings with
spin in the absence of $e-e$ interaction were predicted
\cite{Loss}. With development of fabrication techniques, true
quantum limit of nanoscopic rings containing only a few electrons
was obtained by Lorke and coworkers
\cite{Garcia1997,Lorke1998,Lorke2000}.

On the other hand, carbon based nanostructures, such as
nanoribbons, quantum dots or flakes, and graphene quantum rings
(GQRs), have been the target of intense scrutiny in theory and
experiment since the possibility of controlling their energy
spectrum and hence electronic and novel magnetic properties.
Specially, in recent years, ring-type nanostructures made of
carbon-based materials like carbon nanotubes \cite{Bachtold1999}
and two-dimensional graphene with the large mean free path of
carriers provide new challenges in studying the persistent current
and the AB effect\cite{Russo2008}. Furthermore, signatures of the
AB oscillations in a graphene ring structure for the first time
observed experimentally in a two-terminal setup\cite{Russo2008} by
clear magnetoconductance oscillations with the expected period
corresponding to one magnetic flux quantum  $\phi_0$. In addition,
several studies concerning the AB effect in graphene were
published in theory
\cite{Jackiw2009,Recher,Xavier,Jorg,Abergel2008,Zarenia2010} and
experiment\cite{Huefner2010,Yoo2010,Smirnov2012}.

Recently, based on a noninteracting electron model, the period of
magnetic flux in the AB effect in narrow GQRs with zigzag boundary
conditions has been studied \cite{Romanovsky}. However, graphene
nanoribbons \cite{Fujita,Saffar}, triangular \cite{Potasz2011} and
hexagonal GQRs \cite{Grujic} with zigzag edges are magnetized due
to the $e-e$ interaction. This implies that to study a quantum
interference device based on a graphene nanoring with zigzag
edges, many body effects even in a mean-field approximation must
be included in the theory.

In this paper, the single-band tight-binding (TB) approximation
and the mean-field Hubbard model are combined to investigate the
electronic structures, magnetic properties, and AB oscillations of
hexagonal, triangular and circular narrow graphene rings with
zigzag edges. We show that there are drastic modifications in the
amplitude and the position of AB oscillations due to the e-e
interaction specially near the Fermi energy. We also study the
size effects on the electronic levels and localized magnetic
moments of nanorings for different geometries. Spin-polarized AB
oscillations are observed in the triangular rings in spite of
other ring shapes. We note that an interesting feature of these
Hubbard rings is that, despite being a strongly correlated
electron system, it is easy to introduce physical effects such as
impurity, disorder, magnetic- and electric fields and external
leads in the theory, which are usually much harder to be included
in other models.

The paper is organized as follows: In section II, we present the
single band TB approximation in the presence of $e-e$ interaction
by using the half-filled Hubbard model within the Hartree-Fock
approximation. In Section III, the electronic spectra and magnetic
properties of hexagonal, triangular and circular GQRs, and the
effect of $e-e$ interaction in the AB oscillations will be
illustrated. Finally, we summarize our results in section IV.

\begin{figure}
\centerline{\includegraphics[width=0.72\linewidth]{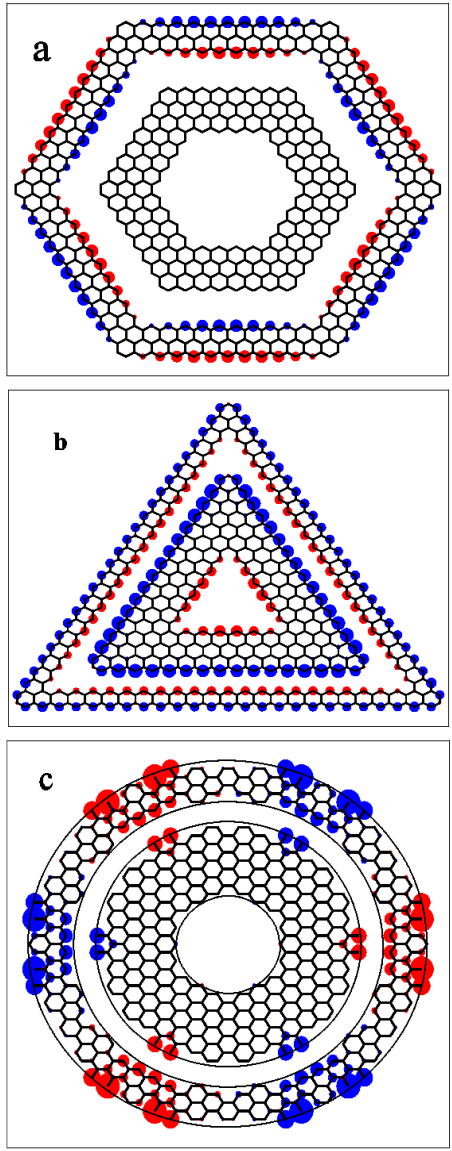}}
\caption{(Color online) Schematic view of (a) hexagonal, (b)
triangular, and (c) circular zigzag-edge graphene rings with
distribution of the local magnetic moments. The blue (red) circles
correspond to the majority (minority) spin electrons. The outer
(narrow) hexagonal and triangular nanorings are described by $W$
and $L$, where $W$ is the number of benzene rings in the thickness
of each ribbon and $L$ is the number of one type of carbon atoms
in the inner edge of each ribbon. In (a), $W=2(3)$ and $L=10(4)$
for the outer (inner) hexagonal ring, (b) $W=3(1)$ and $L= 7(22)$
for the inner (outer) triangular ring, and (c) two circular rings
with different width and size.}
\end{figure}

\section{MODEL AND METHOD}

We simulate the electronic structure of three different GQRs
geometries [shown in Fig. 1] using the $\pi$-orbital tight-binding
model and Hubbard repulsion treated in the mean-field
approximation. This formalism, which includes the $e-e$
interaction in the location of atomic sites, induces localized
magnetic moments on the zigzag-edge atoms. We use the Hubbard
Hamiltonian with a pure AB flux $\phi$. Due to the presence of a
vector potential associated with a uniform magnetic field, the
hopping integrals between the nearest neighbors in the
tight-binding Hamiltonian are modified by a phase factor
\cite{Peierls1933} and as a result, the Hamiltonian for a ring
with $N$ lattice sites can be written as \cite {Fujita}
\begin{equation}
H= \sum_{<i,j>,\sigma}\left(t_{ij} c^{\dagger}_{i,\sigma}c_{j,\sigma}\right)\,\\
+U\sum_{i,\sigma}\langle\hat{n}_{i,-\sigma}\rangle(\hat{n}_{i,\sigma}-{\frac{1}{2}}
\langle \hat{n}_{i,\sigma}\rangle )\  .
\end{equation}

Here, the operator  $c^{\dagger}_{i\sigma}(c_{i\sigma})$ creates
(annihilates) an electron with spin $\sigma$ at site $i$ and
$n_{i\sigma}=c^{\dagger}_{i\sigma}c_{i\sigma}$ is a number
operator. The first term in Eq. (1) describes the hopping of
electrons between neighboring sites (kinetic term) where the
hopping matrix element is defined as
\begin{equation}
t_{ij}=t\exp\left(\frac{ie}{\hbar c}\int_{{\bf r}_i}^{{\bf r}_j}
{{\bf A}({\bf r})\cdot d{\bf s}}\right)\,
\end{equation}
In this equation, ${{\bf r}_i}$ is the position of carbon atom at
site $i$, and ${\bf A}$ is a vector potential in a
symmetric gauge associated with the perpendicular magnetic field
$B$ which can be written as
\begin{equation}
{\bf A}=B{{r_0}^2}/2{(-y/{r^2},x/{r^2},0)}.
\end{equation}
Indeed, the magnetic field $B$ is appliad only in the central
region of absent atoms (antidot) for $r<r_0$ so that the magnetic
flux $\phi$ is $\phi=B\pi({r_0}^2)$ \cite{Romanovsky}. The second
term in Eq.(1) gives the repulsion between electrons occupying the
same site. Accordingly, the magnetic moment at each site of carbon
atoms can be expressed as
\begin{equation}
m_i= \langle {S_i}\rangle=(\langle
\hat{n}_{i,\uparrow}\rangle-\langle
\hat{n}_{i,\downarrow}\rangle)/2 \  .
\end{equation}

Note that, the on-site energy in the tight-binding Hamiltonian is
set to zero, $t = -2.66$ eV is the transfer integral between all
the nearest neighbor sites, and $ U = 2.82$ eV is the Hubbard
parameter which indicates the strength of on-site Coulomb
interactions at each carbon sites of nanorings.

\section{RESULTS AND DISCUSSION }
In order to study the electronic states and magnetic properties of
the nanorings, we start from anti-ferromagnetic configuration as
an initial condition for each structure and solve the mean field
Hubbard Hamiltonian self-consistently. As shown in Fig. 1, we
consider three different types of zigzag-edges rings as hexagonal,
triangular, and circular shapes. In the case of hexagonal and
triangular rings [see Figs. 1(a) and 1(b)], we describe the number
of benzene rings which forms the thickness of each ribbon by $W$
and the number of carbon atoms in the inner edge of each ribbon by
$L$. The hexagonal GQRs consist of six narrow nanoribbons and
therefore have six-fold rotational symmetry, but the triangular
rings with three nanoribbons have three-fold rotational symmetry.
In addition, the circular graphene rings in the present form [see
Fig. 1(c)] have six-fold rotational symmetry. The difference
between the three quantum rings is that both the hexagonal and
triangular rings have well-defined zigzag edges in both inner and
outer edges, while the circular GQRs have both the zigzag and
armchair edges.
\begin{figure}
\centerline{\includegraphics[width=8cm,height=10cm,angle=0]{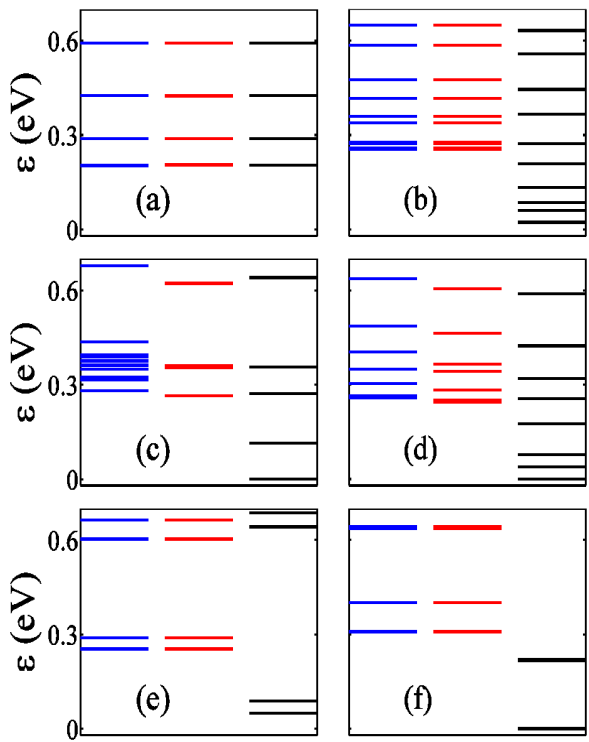}}
\caption{(Color online) Energy levels near the Fermi level for
three different types of zigzag-edges graphene rings. (a) ((b))
hexagonal ring with width W=3(2), (c)((d)) triangular ring with
width W=3(1) and (e)((f)) thicker (thinner) circular ring. In the
figures, the left and middle columns (blue and red lines) show
energy spectrum for majority and minority spins, respectively,
while the right column (black lines) shows energy level in the
absence of electron-electron interaction.}
\end{figure}

In Fig. 1(a), the inner ring, i.e. the thicker structure with $W =
3$ and $L=4$, consists of 288 carbon atoms, while the outer one,
i.e. the thinner structure with $W = 2$ and $L=10$, has 414 carbon
atoms. The hexagonal graphene rings have the same number of A- and
B-type atoms and each edge (inner or outer edge) consists of both
types of atoms. Therefore, according to the Lieb's theorem
\cite{lieb}, the total spin value is zero and each edge has an
antiferromagnetic spin configuration. Furthermore, the
magnetization of each zigzag edge in hexagonal rings strongly
depends on the size and width of the ring. Specially for very
small rings, the edge states on different sides of inner or outer
ring boundaries are subject to strong hybridization, and therefore
spontaneous spin polarization does not occur
\cite{Fernandez,Grujic}. In the case of thicker hexagonal ring
which corresponds to the inner ring in Fig. 1(a), the spin
polarization vanishes completely and a non-magnetic ground state
is obtained. On the other hand, the total spin value for the
thinner hexagonal ring is zero, similar to the thicker hexagonal
graphene ring, while a nonzero spin with maximum value $S=0.12$ is
induced on the inner and outer edges. This means that, due to the
$e-e$ interaction, an antiferromagnetic order forms in this type
of nanorings\cite{Grujic} and the six-fold
rotational symmetry is broken to three-fold symmetry [see the red
and blue circles in Fig. 1(a)].

The two triangular graphene rings with different widths and sizes
have been shown in Fig. 1(b). Both the inner ring (the thicker
structure) with $W =3$ and $L=7$, and the outer one (the thinner
structure) with $W =1$ and $L=22$ have the same 285 carbon atoms.
However, the triangular graphene rings have different number of A-
and B-type atoms and each edge (inner or outer edge) has just one
type of atoms. Therefore, in contrast to the hexagonal GQRs, the
total spin value is non-zero and each edge has a ferromagnetic
spin configuration and therefore, the $e-e$ interaction preserves
the three-fold spatial symmetry. Furthermore, the triangular
graphene rings have opposite spin configurations with different
values of magnetic moments in both edges (inner and outer
edges)\cite{Potasz2011}. The total spin of the thicker (thinner)
structure reaches $S=4.5$ $(1.5)$, which has a maximum value
$S=0.13$ $(0.10)$ in the middle of zigzag edge segment, shown in
Fig. 1(b).

The two symmetric circular graphene rings with six-fold rotational
symmetry and different sizes has been shown in Fig. 1(c). Both
circular rings have 330 atoms. In this type of rings, both zigzag
and armchair edge states exist simultaneously in edge termination.
However, the localized magnetic moments at the armchair edges
almost vanishes. As a result, the magnetization of zigzag edge in
circular GQRs strongly depends on the ring size and width, similar
to the case of hexagonal nanorings. For example, a non-magnetic
phase could be found for a circular ring with width 6 nm and 210
atoms. Although a nonzero magnetization with maximum spin value
$S=0.18$ is produced on the zigzag edges and also in the dangling
bonds, the total spin of our circular rings is zero, due to the
special symmetry between two different sub-lattices. The circular
GQRs have opposite spin configurations in both edges and the
magnetic moment in the inner edges is smaller than the outer
edges, as shown by red and blue circles in Fig. 1(c).

\begin{figure}
\centerline{\includegraphics[width=8.5cm,height=7cm,angle=0]{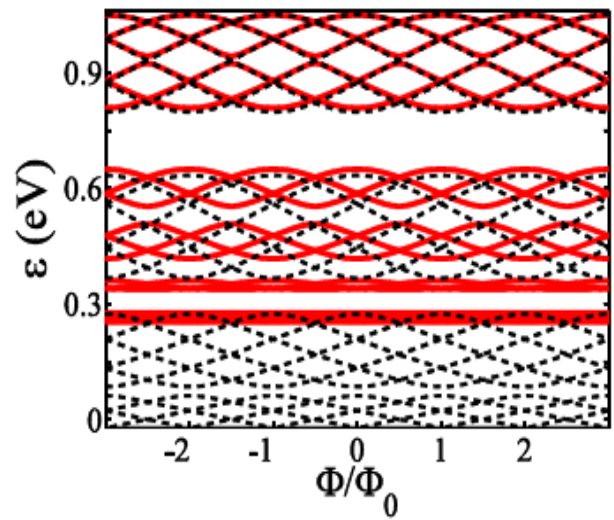}}
\caption{(Color online) Aharonov-Bohm  oscillations of the
hexagonal graphene ring with $W=2$ shown in Fig. 1(a). The solid
(dashed) line shows the AB patterns with (without) considering the
effect of $e-e$ interaction. The oscillations for spin-up and
spin-down electrons are completely degenerate.}
\end{figure}

In Fig. 2 we show how the single particle energy of the various
rings with different sizes varies under the $e-e$ interaction. In
order to compare the interacting and non-interacting systems, the
energy spectra for electrons in  both the systems are plotted near
the Fermi energy that is set to zero. The energy levels for
spin-up and spin-down electrons are also plotted for comparison.
In the case of hexagonal ring with width $W=3$ (288 atoms), the
interacting and non-interacting electrons have the same energy
spectra as shown in Fig. 2(a). However, by increasing the length
of hexagonal ring the effect of $e-e$ interaction increases and
different energy levels, as shown in Fig. 2(b), are produced. Note
that the energy spectra for spin-up and spin-down electrons are
completely degenerate. In all cases, except  the thicker hexagonal
GQR, the $e-e$ interaction causes a significant change in the
distribution of energy levels and induces gap at the Fermi energy.
Although the energy spectrum changes by changing the size of the
rings, the influence of $e-e$ interaction induces different energy
spectra only in the majority and minority spin electrons of the
triangular rings, as shown in Figs. 2(c) and 2(d). By comparing
Figs. 2(a) and 2(b) and Figs. 2(e) and 2(f), one can see that the
effect of interaction on the electronic levels of circular rings
is much similar to the energy levels of hexagonal ones, and no
spin splitting in the electronic levels of these types of
nanorings can be expected.
\begin{figure}
\centerline{\includegraphics[width=7.7cm,height=9cm,angle=0]{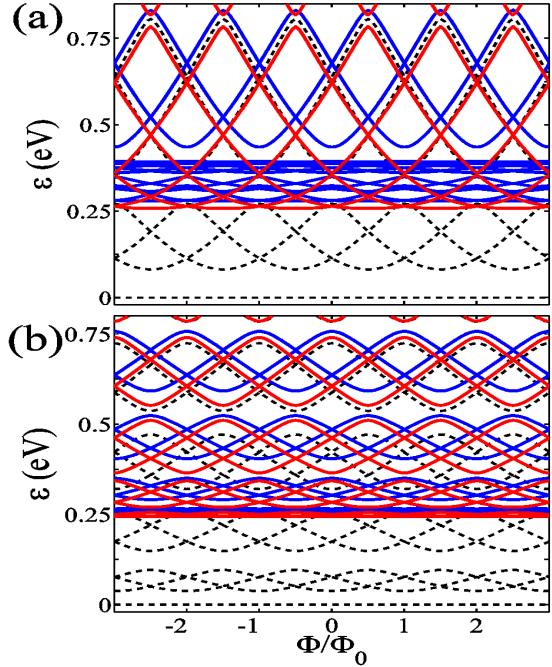}}
\caption{(Color online) AB oscillations as a function of magnetic
flux $\Phi $ for the triangular graphene rings with (a) $W=3$ and
(b) $W=1$ (see Fig. 1(b)). The blue (red) solid lines show AB
pattern for the majority (minority) spin electrons. The dashed
lines show the AB patterns without $e-e$ interaction. Note that
the oscillations for the majority and minority spin electrons are
completely non-degenerate.}

\end{figure}
\begin{figure}
\centerline{\includegraphics[width=7.5cm,height=9cm,angle=0]{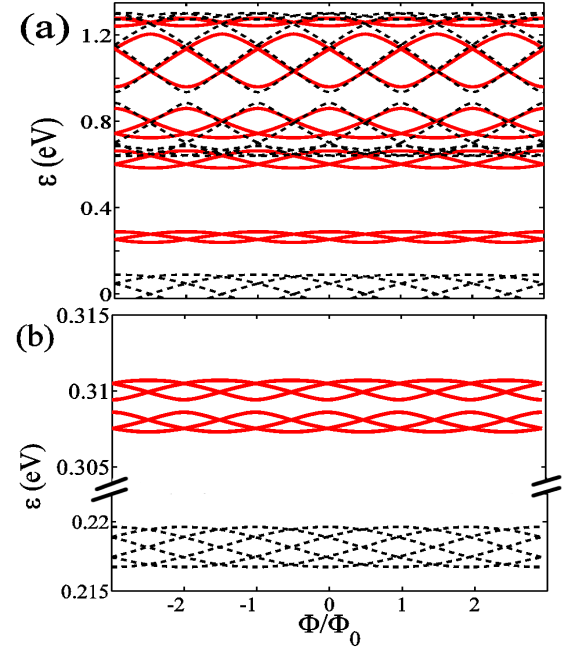}}
\caption{(Color online) AB oscillations as a function of magnetic
flux $\Phi $ for the circular graphene rings shown in Fig. 1(c).
The solid (dashed) lines show AB pattern in the presence (absence)
of $e-e$ interactions for (a) the thicker and (b) the thinner
circular rings. The oscillations for spin-up and spin-down
electrons are completely degenerate.}
\end{figure}

Now, we examine the effect of $e-e$ interaction on the energy
spectra in the presence of uniform magnetic flux threading the
GQRs for various geometries. This interaction does not affect the
energy spectra of the thicker hexagonal ring, i.e., $W=3$ and
$L=4$, as we expect from Fig. 2(a). However, the evolution of AB
oscillations induced by magnetic flux in the thinner hexagonal
ring can be seen in Fig. 3. It is clear that, the period of such
oscillations remains constant, but the energy spectrum and the
amplitude of the oscillations near the Fermi energy are strongly
affected by the interaction. Due to the oscillations, the
energy-gap value between energy levels is flux dependent and the
amplitudes at the lowest energies (around 0.3 eV) are very short,
which means that these states are less efficient in trapping
magnetic flux. In addition, the six-fold rotational symmetry in
the case of non-interacting electrons is broken to a three-fold
rotational symmetry. These changes in the AB oscillations
gradually disappear at higher energies where the change in the
energy gaps decreases and eventually the AB patterns in the case
of interacting and non-interacting electrons become similar. Note
that the AB oscillations for the majority and minority spin
electrons are completely degenerate. Moreover, our calculations
showed that, by increasing the strength of $e-e$ interaction, the
change in the energy gaps at the higher energy levels could also
occur and the amplitude of AB oscillation slightly reduces.

The three energy distributions associated with the triangular
rings as a function of magnetic flux in the case of interacting
and non-interacting electrons have been illustrated in Fig. 4. For
such nanorings, the AB patterns are somewhat complicated and the
AB oscillations of majority and minority spin electrons near the
Fermi energy are fully non-degenerate. A non-zero magnetization,
which comes from magnetic moments, localized on the edge atoms,
causes these spin-polarized oscillations in such triangular rings.
The $e-e$ interaction induces a few energy levels along with very
weak oscillations for both majority and minority spin electrons.
The thinner triangular ring, in the case of interacting and
non-interacting electrons, shows similar behavior at higher
energies and becomes nearly degenerate [see Fig. 4(b)]. With
increasing the width of the ring, however, the net magnetization
increases and the majority and minority spins show different
behavior even at high energy levels [see Fig. 4(a)].

The AB oscillations for the two types of circular nanorings have
been shown in Fig. 5. The main differences between the rings with
interacting and non-interacting electrons fall within an energy
window near the Fermi level. The $e-e$ interaction shifts the
low-lying energy levels to higher energies and slightly decreases
the amplitude of oscillations. In Fig. 5(a), the energy gap
between AB oscillations increases at integer and half-integer
values of $\phi$, depending on the energy eigenvalues. Moreover,
in the case of thinner ring, the amplitude of oscillations, which
have been plotted in a small range of energies, is considerably
weaker than that of the thicker one [see Fig. 5(b)].

The obtained results indicate that the $e-e$ interaction, which
induces localized magnetic moments on the edges of GQRs, is able
to reduce the magnetic symmetries compared to the structural
symmetries and affects the AB oscillations dramatically, relative
to the nanorings with non-interacting electrons. Note that this
interaction cannot induce localized magnetic moments in the
armchair nanorings and therefore, the AB oscillations are not
affected by this kind of interaction.

\section{Conclusion}
In summary, the influence of $e-e$ interaction on the AB pattern
in zigzag-edge GQRs was studied by means of single-band
tight-binding Hamiltonian and the mean-field Hubbard model. This
interaction induces localized magnetic moments on the the inner
and outer edges of graphene nanorings which are strongly size and
shape dependent. In addition, the interaction modifies the AB
oscillations by changing energy levels near the Fermi energy and
hence, a reduction in the amplitude of oscillations can be seen in
all nanoring structures. The change in the amplitude and the
position of AB oscillations is different from one nanoring to the
other.

In the case of hexagonal and circular rings, the six-fold
rotational symmetry changes to the three-fold symmetry due to the
distribution of localized magnetic moments, while the degeneracy
between the spin-up and spin-down electrons is not broken which
causes unpolarized AB oscillations. On the other hand, due to the
non-zero magnetic moment in the triangular rings, the degeneracy
between the majority and minority of electrons is broken;
therefore, spin-polarized AB oscillations can be expected in this
type of nanorings. This polarization could provide us a
possibility to generate the persistent spin current (SC)
intrinsically in the zigzag-edge nanorings which are obtained by
cutting and patterning the graphene sheets into the nanorings with
special sizes and geometries. Thus, as well as being of
fundamental interest, systematic experimental studies of magnetic
edge states in zero-dimensional carbon structures like graphene
nanorings might be useful for future spintronic applications.

\section*{ACKNOWLEDGMENTS}
This work financially supported by university of Kashan
under the Grant No. 228762.

\end{document}